# Screening of Fungi for the Application of Self-Healing Concrete

Rakenth R. Menon[1], Jing Luo[2], Xiaobo Chen[3], Hui Zhou[3], Zhiyong Liu[4], Guangwen Zhou[1,3], Ning Zhang[2,5*], Congrui Jin[1,3*]

*1 Department of Mechanical Engineering, Binghamton University, NY 13902, USA*

*2 Department of Plant Biology, Rutgers University, New Brunswick, NJ 08901, USA*

*3 Materials Science and Engineering Program, Binghamton University, NY 13902, USA*

*4 Center for Neurodegeneration and Experimental Therapeutics, Department of Neurology, University of Alabama at Birmingham, Birmingham, AL 35294, USA*

*5 Department of Biochemistry and Microbiology, Rutgers University, New Brunswick, NJ 08901, USA*

*Corresponding author:* ningz@rutgers.edu; cjin@binghamton.edu

**Abstract**

Concrete is susceptible to cracking owing to drying shrinkage, freeze-thaw cycles, delayed ettringite formation, reinforcement corrosion, creep and fatigue, etc. Continuous inspection and maintenance of concrete infrastructure require onerous labor and high costs. If the damaging cracks can heal by themselves without any human interference or intervention, that could be of great attraction. In this study, a novel self-healing approach is investigated, in which fungi are applied to heal cracks in concrete by promoting calcium carbonate precipitation. The goal of this investigation is to discover the most appropriate species of fungi for the application of biogenic crack repair. Our results showed that, despite the significant pH increase owing to the leaching of calcium hydroxide from concrete, *Aspergillus nidulans* (MAD1445), a pH regulatory mutant, could grow on concrete plates and promote calcium carbonate precipitation.

## 1. Introduction

Sustainable infrastructure is the key to creating a sustainable community, due to its significant impact on energy consumption, land use, and global economy. However, many countries are facing the downfall of progressively aging infrastructure that needs rehabilitation. In particular, concrete infrastructure suffers from serious deterioration owing to the effect of various physical and chemical phenomena, such as drying shrinkage, freeze-thaw cycles, reinforcement corrosion, creep and fatigue, and delayed ettringite formation, all of which could lead to concrete cracking.

Cracks themselves may not significantly reduce the load-carrying capacity of concrete in the short run, but they considerably weaken the durability of concrete structures, as they channel in water, oxygen, and carbon dioxide, which could potentially corrode the steel reinforcement. Moreover, cracking may promote severe degradation of the non-mechanical properties of concrete, such as the radiation-shielding properties of concrete elements used in nuclear applications. Nowadays, concrete has been the key construction material for reactor containment and biological shielding structures, which are essential components of the nuclear reactors in service worldwide for power generation. In addition, cementitious grouts, mortars, and concrete are also often used to provide shielding and encapsulation of various radioactive waste materials from military, research, and power generation applications. Some waste isotopes as well as their decay products will become a serious radiation hazard for hundreds of thousands of years, which requests exceptionally durable storage.

In view of the remarkable social significance of concrete infrastructures and their exceptional service demands, the maintenance and inspection for concrete structures have come into focus. However, continuous inspection and maintenance usually require onerous labor and high investments, which presents a huge and costly challenge. Fortunately, inspired by the amazing capability of the human body to fix broken bones through mineralization, researchers have conducted numerous investigations to equip concrete structures with self-healing properties and generated many innovative solutions[1-23].

So far, concrete can fix its own cracks mainly through the following three mechanisms: autogenous healing, embedment of polymeric material, and bacteria-mediated $CaCO_3$ precipitation, which have been summarized in a comprehensive review provided by Seifan et al.[20] Among them, the biotechnological approach by using mineral-producing microorganisms is most desirable. It has been demonstrated that in the environment of concrete, $CaCO_3$ can be precipitated by certain species of bacteria through biologically induced mineralization processes[1-21]. Owing to its sufficient compatibility with concrete compositions, $CaCO_3$ is often regarded as the most appropriate filler for concrete cracks.

Bacteria can precipitate $CaCO_3$ mainly through three different pathways, as summarized below. (1) Certain species of ureolytic bacteria have been explored based on their ability to indirectly cause the formation of calcium precipitates by increasing the local pH via catalyzing the hydrolysis of urea to ammonium[1-5]. (2) Microbiologists at Delft University of Technology in the Netherlands[11-13] tested the second pathway, in which aerobic oxidation of organic acids results in production of $CO_2$, leading to $CaCO_3$ production in the environment of concrete. However, in the case of low oxygen concentration, such as for most underground structures, the efficiency of this approach can be limited. (3) The third pathway is known as dissimilatory nitrate reduction[16]. This approach is less efficient than the other two in terms of the amount of produced $CaCO_3$, but it can be applied in anaerobic zones.

## 2 Fungi-Mediated Self-Healing Concrete

Although the research on bacteria-mediated self-healing concrete indeed achieved a certain level of success, it still suffers from serious limitations. So far, the viability of bacterial spores embedded in concrete is generally less than six months[13], which is far from being practical considering the fact that the lifetime of concrete infrastructure can easily be fifty years or even a century. The detrimental environment of concrete, such as very high pH values, tiny pores, serious moisture deficit, varied temperatures, and limited nutrient availability, dramatically influences the microbial metabolic activities and makes bacteria and their spores susceptible to death. In addition, owing to the limited ability of bacteria to produce large amounts of $CaCO_3$, bacteria can only heal small cracks with crack widths less than 0.8 mm[10-20].

To address the above-mentioned problems, further investigation on alternative microorganisms becomes potentially essential. In fact, microorganisms are very diverse, and besides bacteria, they also include Archaea, protists, and fungi. Among them, fungi form a species-rich group of eukaryotic organisms, which exhibits vast biodiversity with more than 100 K described species and approximately 1.5 M unknown species. Common fungi include yeasts, lichen-forming fungi, molds, and mushrooms, etc.

Recent studies in geomycology demonstrated that certain species of fungi can play a central role in $CaCO_3$ precipitation[24,25], but those species have never been tested in the environment of concrete. Currently, the investigation of fungi has been mostly focusing on their importance in the degradation of organic matter, and their connection with inorganic constituents is limited to acquisition of mineral nutrition by mycorrhizal fungi as well as mineral weathering of lichens and ectomycorrhizae.

The cells of filamentous fungi grow as threadlike structures called hyphae, which possess multiple nuclei and grow apically with new apices emerging from the formation of lateral branches, creating an intertwined 3D network called mycelium. In fact, the existing literature suggests that filamentous fungi have unique features to be used in various applications of biomineralization-based technologies[26-30]. For example, compared with other microbial groups, filamentous fungi exhibit higher surface-to-volume ratios, and therefore possess a larger fraction of organic substrates available for mineral precipitation.

The calcification of fungal filaments is a complex process, and it remains incompletely understood[29-36]. However, it has been concluded that there are two critical factors determining the amount of $CaCO_3$ production, i.e., carbonate alkalinity and $Ca^{2+}$ concentration[24]. The metabolic activities of filamentous fungi that can increase carbonate alkalinity typically include water consumption, physicochemical degassing of fungal respired $CO_2$, oxidation of organic acids, nitrate assimilation, and urea mineralization[24]. In addition to carbonate alkalinity, fungal metabolic activities can also affect calcium concentration. $Ca^{2+}$ concentration within fungal cells should be strictly controlled, i.e., most of the calcium ions need to be gathered at the apex for apical growth and sharply reduced in subapical regions[24]. To maintain this sharp gradient, fungi need to effectively regulate $Ca^{2+}$. $Ca^{2+}$ in the cytoplasm is kept at sufficiently low concentrations by deliberately pumping it out of the cell or by binding it onto cytoplasmic proteins[24].

In addition, fungi can also influence calcium concentration out of self-protection. Metals that are required for fungal growth and metabolism may become toxic if their concentrations are too high. However, unlike other microbial group, many species of fungi can survive and prosper in regions that are seriously polluted by metals. Precipitation of metal minerals onto hyphae was considered as one of the most important mechanisms to explain the superior metal tolerance in fungi. Concrete is a calcium-rich environment, which represents a stress for fungal cells owing to calcium cytotoxicity and subsequent osmotic pressure[24]. The production of calcium oxalates has been regarded as a method to decrease their internal $Ca^{2+}$ concentration[24]. Precipitation of $CaCO_3$ may be due to a similar passive mechanism to immobilize excessive $Ca^{2+}$. Excessive alkalinity could also pose a source of stress and precipitation of $CaCO_3$ may be due to intracellular protection[24].

While bacteria promote mineral precipitation only through induced biomineralization processes, fungi can do it through both induced biomineralization and organomineralization processes. There exists a substance called chitin in fungal cell walls, which is a long carbohydrate polymer forming a substrate that could considerably reduce the required activation energies for nuclei formation so that the interfacial energy between the fungus and the mineral crystal becomes significantly lower than the one between the mineral crystal and the solution[32,33]. Thus, both living and dead fungal biomass can bind ions onto their cell walls, resulting in nucleation and deposition of mineral phases[31]. Bound $Ca^{2+}$ can then interact with the dissolved carbonate, resulting in $CaCO_3$ precipitation on the fungal hyphae.

Therefore, thanks to their ability to directly and indirectly promote $CaCO_3$ production, fungi could be used as self-healing agents. The goal of this investigation is to discover the most appropriate species of fungi for the application of biogenic crack repair. To make fungi-mediated self-healing concrete, the fungal spores alongside their nutrients, will be blended into the concrete before the curing process starts. When cracks appear and water trickles into the concrete, the dormant fungal spores will wake up, grow, consume the nutrient soup, and promote $CaCO_3$ precipitates to fix the cracks *in situ*. After the cracks are finally healed, the bacteria or fungi will make spores and go dormant once more – ready to start a new cycle of self-healing when cracks form again. For existing concrete infrastructures with cracks, the fungal spores and their nutrients can be injected or sprayed into the cracks.

Besides wild-type fungal strains, genetically engineered fungi are also important candidates for self-healing concrete. Since the extremely high pH of the concrete environment can be handled by only a few fungi, pH regulatory mutants were the focus here. It is well known that many microorganisms that are capable of growing over a relatively wide pH range can adjust their gene expression according to the ambient pH levels[37,38]. Fungi make responses to the extracellular pH by means of activation of a dedicated transcription factor, PacC[39-44]. Mutants in the *pacC* regulatory gene are extremely heterogeneous in phenotype[37]. A major class of mutations are gain-of-function *pacC$^c$* created by truncating the C-terminal region[38]. The *pacC$^c$* mutations bypassing the need for the extracellular pH signal result in permanent activation of alkaline genes and superrepression of acidic genes, which leads to alkalinity mimicry[41]. Irrespective of ambient pH, fungi exhibiting alkalinity-mimicking mutations believe that they are always at alkaline pH and trigger a gene expression pattern similar to that of the wild-type grown at high pH values, which is exactly what is needed for self-healing concrete.

## 3. Materials and Methods

In our previous investigation[45], we have found that the spores of a wild-type *Trichoderma reesei* (ATCC13631) germinated into hyphal mycelium on concrete plates and grew well. However, there exist only a few studies on the pH regulation of *T. reesei*. In filamentous fungi, the best characterized member for gene regulation by ambient pH is *Aspergillus nidulans*[38]. Therefore, gene manipulations are easily achievable in *A. nidulans*. In this study, three different types of alkalinity-mimicking mutants of *A. nidulans,* i.e., MAD1445, MAD0305, and MAD0306, will be tested.

The following five different wild-type strains were purchased from American Type Culture Collection (ATCC): *Rhizopus oryzae* (ATCC22961), *Phanerochaete chrysosporium* (ATCC24725), *A. nidulans* (ATCC38163), *A. terreus* (ATCC1012), and *A. oryzae* (ATCC1011). The genomes of these fungal strains have been sequenced and annotated and are publicly available[46-54]. In addition, these species do not exhibit any toxicity and belong to Biosafety Level 1 (BSL-1), which is the lowest risk level.

The above-mentioned fungi are all filamentous fungi. In addition to filamentous fungi, there are also single-celled fungi, i.e., yeasts, that do not form hyphae. One of the most well-known species of yeast is *Saccharomyces cerevisiae*, also called baker's yeast, as it has been instrumental to winemaking, baking, and brewing since ancient times. Recently, a series of publications were released on the success of designing synthetic set of *S. cerevisiae* chromosomes, which could allow *S. cerevisiae* to rapidly evolve and rearrange its genome, optimizing itself for certain applications[55-57]. Thus, in this study, *S. cerevisiae* will also be tested.

To characterize the fungal precipitates, X-ray diffraction (XRD), scanning electron microscope (SEM), and transmission electron microscope (TEM) were applied in this study. XRD is an established technique to identify unknown crystalline phases[58,59]. SEM has been used to visualize fungal precipitates[60,61], and TEM has been used to study fungal-biotite interfaces and weathered alkali feldspars[62,63]. SEM will be applied to analyze the morphology and composition of the fungal precipitates, which will complement the very local characterization by TEM.

In this section, the experimental procedures of gene replacement to obtain alkalinity-mimicking mutations, survival test of fungi on concrete plates, as well as phase identification and microscopic characterization of fungal precipitates will be presented. The procedures of the yeast strain identification and preparation of concrete specimens are described in the Supplementary Information.

### 3.1 Gene Replacement Procedure to Obtain Alkalinity-Mimicking Mutations

The full genotypes of the strains investigated in the current work are given in Table 1. The detailed about the parental strains and the selection procedures of isolating the *pacC* mutations were previously published[64-68].

*Table 1. The full genotypes of the strains.*

| Strain | Genotype | Reference[64-68] |
|---|---|---|
| MAD1445 | *yA2 pabaA1 pacC14900* | Penas et al. 2007; Hervas-Aguilar et al. 2007 |
| MAD0305 | *pabaA1 pacC14* | Caddick et al., 1986; Tilburn et al. 1995; Mingot et al. 1999 |
| MAD0306 | *pabaA1 pacC202* | Tilburn et al. 1995; Mingot et al. 1999 |

Briefly, the gene replacement procedure to obtain alleles *pacC14* and *pacC202* can be described as follows. The parental strains with genotype *pabaA1 gatA2 palF15 fwA1* (*p*-aminobenzoate-requiring, lacking ω-amino acid transaminase, mimicking growth at acidic pH, and having fawn conidial color) were mutagenized with ultraviolet treatment followed by growth selection on media supplemented with 5 mM γ-amino-*n*-butyrate in presence of 10 mM ammonium tartrate as nitrogen source[69-71]. The *pacC14* mutant strain was selected as it mimics the growth at alkaline pH values, downregulates the γ-amino-*n*-butyrate permease, decreases the γ-amino-*n*-butyrate level in cells, and thus relieves the γ-amino-*n*-butyrate-induced cytotoxicity. The *pacC202* strain was a spontaneous mutant strain derived from the parental strains with genotype *pabaA1 sasA60* (*p*-aminobenzoate-requiring and semialdehyde sensitive), selected as it relieves the toxicity of 10 mM γ-amino-*n*-butyrate. The *pacC* genomic sequences in *pacC14* and *pacC202* strains were then obtained by Polymerase Chain Reaction (PCR) amplification.

The procedure to obtain MYC$_3$-tagged wild-type *pacC* allele, designated as *pacC900*, as well as the engineered allele *pacC14900* can be described as follows. A null allele of *pacC* was first constructed by transforming the orthologous *pyr4* (orotidine 5'-phosphate decarboxylase) gene of *Neurospora crassa* flanked by 750 bp of upstream and 700 bp of downstream *pacC* genomic sequences into a *pyrG89* (resulting in pyrimidine auxotrophy) *pacC14* strain. Transformation was carried out using linearized fragments containing the *pacC* moiety of the plasmid. The long flanking sequence ensures high efficiency of homologous recombination-mediated gene replacement. The transformants displaying pyrimidine prototrophy were then selected and tested by Southern blot analyses to confirm that *pacC14* was replaced by *pyr4*. The *pacC* null strain was then used as recipient strain and transformed with linearized MYC$_3$-tagged *pacC* carrying C2437A mutation flanked with genomic sequence 825 bp upstream of *pacC* start codon and 925 bp downstream of stop codon[72]. Transformants displaying *pacC* colony morphology were tested for pyrimidine auxotrophy. The gene replacement was again confirmed using Southern blot analyses. The genotypes in the resulting *pacC14900* strain were further confirmed by PCR sequencing. The protein expression was confirmed by Western blot analyses using anti-MYC antibody.

## 3.2 Survival Test of Fungi on Concrete Plates

In this study, potato dextrose agar (PDA) was chosen as growth medium. To investigate the impact of the extracellular pH on fungal growth, the inert pH buffer 3-(N-morpholino)propanesulfonic acid (MOPS, 20mM, pH 7.0) was added to some of the growth media. Sterilization of concrete plate was carried out in a steam autoclave at 121 °C under 15 psi of pressure for 30 min. A 5 mm diameter mycelial disc of each fungal strain removed from 7-day-old cultures was aseptically placed at the center of each 60 mm Petri dish containing 10 ml growth medium. In some Petri dishes, there exists a concrete plate of 3 mm thickness at the bottom. Sterile PDA plugs were used as the negative inoculum control. The Petri dishes were incubated at 22 °C and 30 °C, respectively, for 21 days.

To summarize, as shown in Fig. S1 in the Supplementary Information, for each type of fungal strain, eight different cases were investigated including PDA30, PDA22, MPDA30, MPDA22, CPDA30, CPDA22, CMPDA30, and CMPDA22. All the cases were investigated independently in triplicates.

Radial growth was measured based on two perpendicular diameters. The fungal samples prepared by the tape touch method[73] were studied using an optical microscope. pH values were recorded after 21 days of incubation by five independent measurements using an Orion 9110DJWP double junction pH electrode, which is a semi-micro combination pH electrode.

## 3.3. Phase Identification of Fungal Precipitates

Phase identification of the fungal precipitates were performed on a Siemens-Bruker D5000 powder diffractometer with Cu-Kα radiation in the θ/θ configuration operating at 40 kV and 30 mA. The samples were analyzed over the range of 10° to 80° 2θ at a scan rate of 1°/min with an increment of 0.02° 2θ. Before the measurements, the solid precipitates were isolated by dissolving the fungal sample in a vat of dilute bleach (NaOCl less than 3% by weight) at room temperature and washing it repeatedly in methanol following the published protocol[59].

## 3.4 Microscopic Characterization of Fungal Precipitates

### 3.4.1 Scanning Electron Microscope (SEM)

The fungal precipitates were studied by a Zeiss Supra 55 VP Field Emission SEM with an EDAX Genesis energy-dispersive X-ray spectrometer (EDS) at accelerating voltages of 5 kV to 20 kV. The collected solids were placed in an oven at 50 °C for two days, and then were sputter-coated with carbon by an Emitech K950X Carbon Vacuum Evaporator in order to ensure electrical conductivity. The chemical microanalysis technique EDS was applied in conjunction with the SEM analysis.

### 3.4.2 Transmission Electron Microscope (TEM)

TEM analysis was performed by a JEOL JEM-2100F TEM operating at 200 kV. The precipitates were collected by centrifugation (6000×g, 10 min) and washed multiple times in methanol to get rid of the growth medium. The particles were dried at room temperature then ground using a set of agate mortar and pestle into fine powders. The powders were then collected on copper grids overlain with a carbon-coated collodion film prior to the examination. Bright-field imaging (BFI), selected area electron diffraction (SAED), and high-resolution transmission electron microscopy (HRTEM) were applied to analyze the calcite phases.

### 3.5 Pore Size Distribution Inside Cement Paste Specimen

High-resolution X-ray computed microtomography (μCT) was applied to characterize the pore size distribution inside the cement paste specimen. Images were acquired using UltraTom X-ray μ-CT system equipped with a LaB6 X-ray source and a CCD camera with resolution of 4000 × 2624 pixels. Small cubic samples of 0.4 cm × 0.4 cm × 0.4 cm were used, for which the voxel size of the obtained images will be about 1.2 μm. Samples were firmly fixed on a cylindrical holder mounted on a rotating table using an adhesive paste. Acquisition parameters were set as follows: source 230 kV, voltage 90 kV, intensity 200-300 μA, filament LaB6, frame rate 0.4, and 8 averaging images. To impose small angular increments to cover 360° along the rotation axis, a total of 1664 phase contrast images were collected. The acquisition duration was about 30 min for each sample. Based on the raw data measured by μCT, the reconstruction of inner microstructure was carried out by using the Octopus Imaging software[74]. Images were post-processed by using the ImageJ software[75].

## 4. Results and Discussion

### 4.1 Mutant Sequence Changes and Resulting Proteins

The mutant sequence changes in *pacC* and the resulting proteins are summarized in Table 2. Note that c denotes constitutive phenotype, which bypasses the requirement for the pH signal and mimics alkaline growth conditions; and an asterisk denotes a stop codon. It can be seen that the *pacC14* genotype carries a C2437A allele, causing an early stop codon that results in a C-terminal truncation of PacC protein at 492 amino acid. The *pacC202* genotype has a deletion from 2353 bp to 2555 bp, resulting in an open reading frame shift after amino acid 464 and an early stop codon. The resulting *pacC14900* strain has a MYC$_3$-tagged *pacC* carrying the *pacC14* mutation at the *pacC* locus.

*Table 2. Mutant sequence changes in pacC and resulting proteins.*

| Genotype | Phenotype | Mutation | Mutant protein | Reference[65,66] |
|---|---|---|---|---|
| *pacC14* | c | C2437A | 1→492* | Tilburn et al. 1995 |
| *pacC202* | c | Δ2353→2555 | 1→464 IDRPGSPFRISGRG* | Tilburn et al. 1995 |
| *pacC14900* | c | C2437A | 1→492* | Hervas-Aguilar et al. 2007 |

### 4.2 Identification of the Yeast Strain

The yeast strain has 99% ITS sequence similarity to the *S. cerevisiae* ex-type culture CBS 1171 (NR_111007) and has the same morphological features of *S. cerevisiae*. Therefore, it is confirmed as the baker's yeast, i.e., *S. cerevisiae*.

## 4.3 Fungal Growth on Concrete Plates

The measurement results show that, because of the leaching of calcium hydroxide from concrete, the pH values increased significantly from 6.5 to 13.0 (Table 3). On the concrete plates, only one type of pH regulatory mutants of *A. nidulans*, i.e., MAD1445, has been found to be able to grow well (Table 4). At 30 °C, its growth rate was 3.2 mm/day in the case of CMPDA30. Plentiful conidia were found on the concrete plates, which possess similar morphology with those found in the Petri dishes without concrete (Fig. 1). However, *A. nidulans* (MAD1445) did not grow in the cases of CPDA22 or CPDA30. The other eight species grew well in all the cases without concrete but did not grow on any concrete plate. The fungal growth in each case and the associated optical microscopic images are shown in Fig. S1, Fig. S2, and Fig. S3 in the Supplementary Information.

Note that of all these three strains, which carry *pacC* constitutive mutations and otherwise possess very similar phenotypes, only MAD1445 grows well on the concrete plates in the presence of MOPS. This could be due to the fact that other unknown genetic factors could lead to increased survival against high pH values. The three strains were obtained through either mutagenesis or homologous recombination. Although the *pacC* locus was sequence-verified, it is possible that mutations happened at other undesired loci of the MAD1445 strain resulted in increased survival in the alkaline environment.

*Table 3. pH measurement results after 12 days of inoculation.*

|  | PDA | MPDA | CPDA | CMPDA |
|---|---|---|---|---|
| Control | 6.13±0.12 | 6.83±0.14 | 13.14±0.14 | 12.02±0.11 |
| *Aspergillus nidulans* (ATCC38163) | 6.50±0.12 | 7.22±0.12 | 13.04±0.13 | 11.60±0.10 |
| *Aspergillus nidulans* (MAD1445) | 6.42±0.09 | 7.12±0.15 | 13.22±0.16 | 11.21±0.10 |
| *Aspergillus nidulans* (MAD0305) | 6.31±0.12 | 7.08±0.13 | 13.01±0.10 | 12.11±0.10 |
| *Aspergillus nidulans* (MAD0306) | 6.19±0.19 | 7.11±0.19 | 13.11±0.06 | 11.32±0.08 |
| *Aspergillus oryzae* (ATCC1011) | 6.34±0.13 | 7.14±0.20 | 13.63±0.13 | 12.19±0.19 |
| *Aspergillus terreus* (ATCC1012) | 6.33±0.22 | 7.03±0.15 | 13.14±0.20 | 13.04±0.08 |
| *Rhizopus oryzae* (ATCC22961) | 6.71±0.20 | 7.02±0.06 | 13.24±0.19 | 12.84±0.12 |
| *Phanerochaete chrysosporium* (ATCC24725) | 6.81±0.20 | 7.00±0.05 | 13.31±0.13 | 12.11±0.13 |
| *Saccharomyces cerevisiae* (Yeast) | 6.25±0.13 | 7.04±0.19 | 13.01±0.06 | 12.81±0.15 |

*Table 4. Average growth rates (mm/day) after 21 days of inoculation (n = 6). Note that [mean ± standard deviation] followed by different letters are remarkably different according to Tukey's HSD test (p = 0.05).*

|  | PDA30 | PDA22 | MPDA30 | MPDA22 | CPDA30 | CPDA22 | CMPDA30 | CMPDA22 |
|---|---|---|---|---|---|---|---|---|
| *Aspergillus* | 2.6±0. | 2.6±0. | 0.5±0.12[b] | 0.7±0.23[b] | 0 | 0 | 0 | 0 |
| *nidulans* (ATCC38163) | 07[a] | 25[a] | | | | | | |
| *Aspergillus nidulans* (MAD1445) | 3.9±0.18[a] | 1.9±0.14[c] | 1.2±0.15[d] | 1.2±0.17[d] | 0 | 0 | 3.2±0.52[b] | 0.9±0.23[e] |
| *Aspergillus nidulans* (MAD0305) | 3.2±0.17[a] | 2.1±0.06[b] | 1.6±0.08[c] | 1.5±0.21[c] | 0 | 0 | 0 | 0 |
| *Aspergillus nidulans* (MAD0306) | 2.1±0.20[a] | 1.4±0.33[b] | 0.3±0.06[c] | 0.3±0.12[c] | 0 | 0 | 0 | 0 |
| *Aspergillus oryzae* (ATCC1011) | 4.3±0.06[a] | 2.8±0.14[c] | 3.3±0.09[b] | 2.3±0.13[d] | 0 | 0 | 0 | 0 |
| *Aspergillus terreus* (ATCC1012) | 4.6±0.46[a] | 2.8±0.35[b] | 1.8±0.06[c] | 1.7±0.22[c] | 0 | 0 | 0 | 0 |
| *Rhizopus oryzae* (ATCC22961) | 7.8±0.07[a] | 7.8±0.15[a] | 7.8±0.06[a] | 7.8±0.12[a] | 0 | 0 | 0 | 0 |
| *Phanerochaete chrysosporium* (ATCC24725) | 7.8±0.06[a] | 7.8±0.07[a] | 7.8±0.15[a] | 4.2±0.07[b] | 0 | 0 | 0 | 0 |
| *Saccharomyces cerevisiae* (Yeast) | 1.9±0.09[a] | 1.9±0.23[a] | 1.9±0.29[a] | 1.5±0.11[b] | 0 | 0 | 0 | 0 |

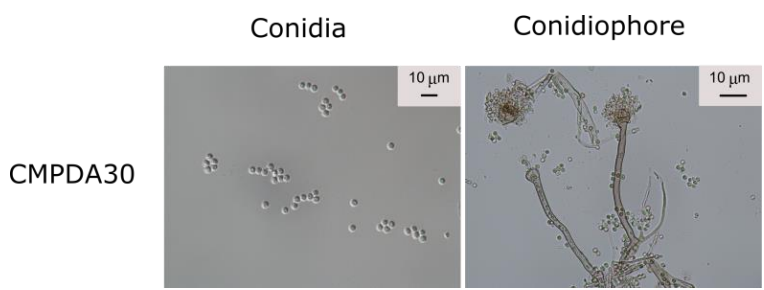

*Figure 1. For the case of A. nidulans (MAD1445), plentiful conidia were found in the case of CMPDA 30. The diameter of A. nidulans spores is typically larger than 3 μm. More images are shown in Fig. S3 in the Supplementary Information.*

### 4.4 Characteristics of the Fungal Precipitates

The XRD analysis shows that the precipitates promoted by fungi are indeed $CaCO_3$ crystals (Fig. 2). The sharp peak at approximately 30° 2θ indicates highly crystalline phases of calcite. In comparison, the mortar specimens obtained from the control medium were composed of both quartz and calcite.

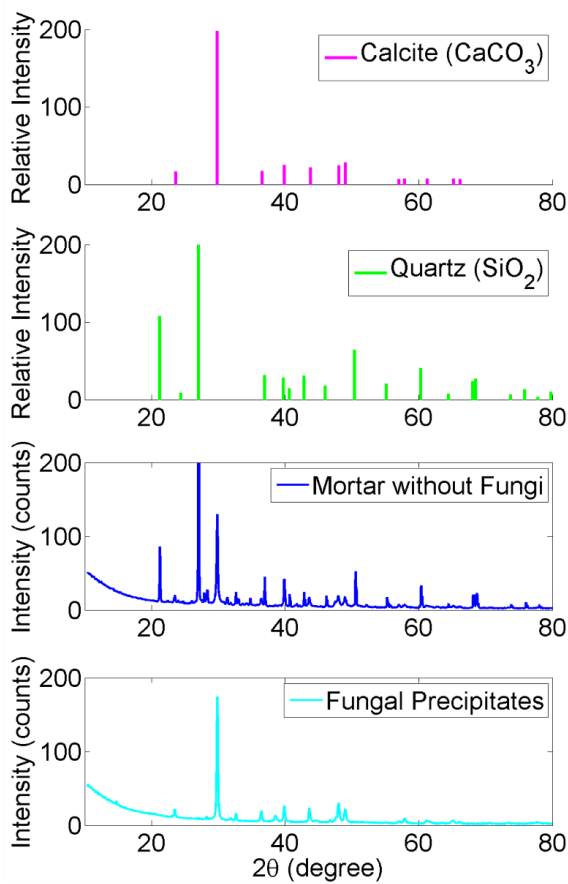

*Figure 2. XRD results for the solids collected from A. nidulans (MAD1445)-inoculated media and fungus-free media.*

Fig. 3(a) shows massive mineral crystals in the *A. nidulans* (MAD1445)-inoculated media: CMPDA22 and CMPDA30. The particles had regular shapes with well-defined crystalline faces, indicating high crystallinity of the precipitates. The mineral crystals exhibited a range of crystalline morphologies from a single pure bulk crystal to plate-like crystals stacked upon each other. Besides, the crystals confirmed fungal involvement, as many cylindrical holes were found in the crystals, which probably occurred in the space inhabited by fungi. The holes also suggested that fungal filaments functioned as nucleation sites during the biomineralization process. As shown in Fig. 3(b), EDS results show that the crystals are composed of calcium, carbon, and oxygen and their atomic percentage approximately matches that of $CaCO_3$. Note that EDS data may carry error if the sample is not homogeneous or flat. In our test, EDS data are obtained from several relatively flat sample areas, and the results are consistent. In comparison, the amount of calcium precipitation in the control medium was significantly reduced, as evidenced by the absence of faceted particles (Fig. 3(c)).

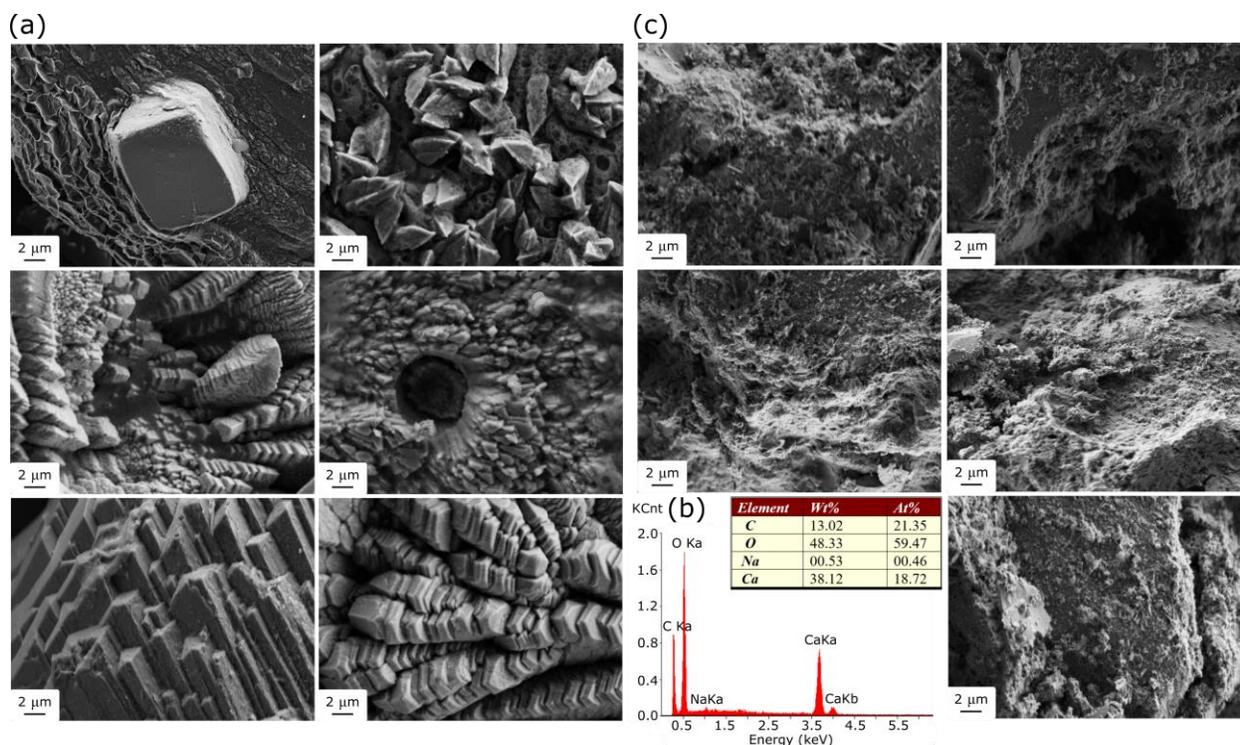

*Figure 3. SEM and EDS results of the collected solids: (a) A. nidulans (MAD1445)-inoculated media; (b) EDS spectra; and (c) fungus-free medium.*

TEM analyses, as shown in Fig. 4, further confirm the crystalline nature of the $CaCO_3$. Fig. 4(a) shows a bright-field TEM image of the fragments. Fig. 4(b) is a SAED pattern from the collection of the particles shown in Fig. 4(a), which demonstrates the crystalline nature of the precipitated particles, as evidenced by the presence of well resolved diffraction rings of (104), (110), (11-6), (122), and (1010) planes, matching well with the calcite phase. Fig. 4(c) is a higher magnification TEM image of the precipitate particle, which shows the presence of crystalline lattice (moiré fringes are also visible due to the overlapping of particles). Fig. 4(d) is a HRTEM image obtained from the area as marked with the black dashed square in Fig. 4(c). As shown in Fig 4(e), the interplanar spacing of the crystalline planes in Fig. 4(d) is measured to be 0.308 nm, which matches with the interplanar spacing of (104) of the hexagonal crystal of calcite with the lattice parameters of a=b=4.989 Å, c=17.062 Å, α=β= 90°, and γ=120°.

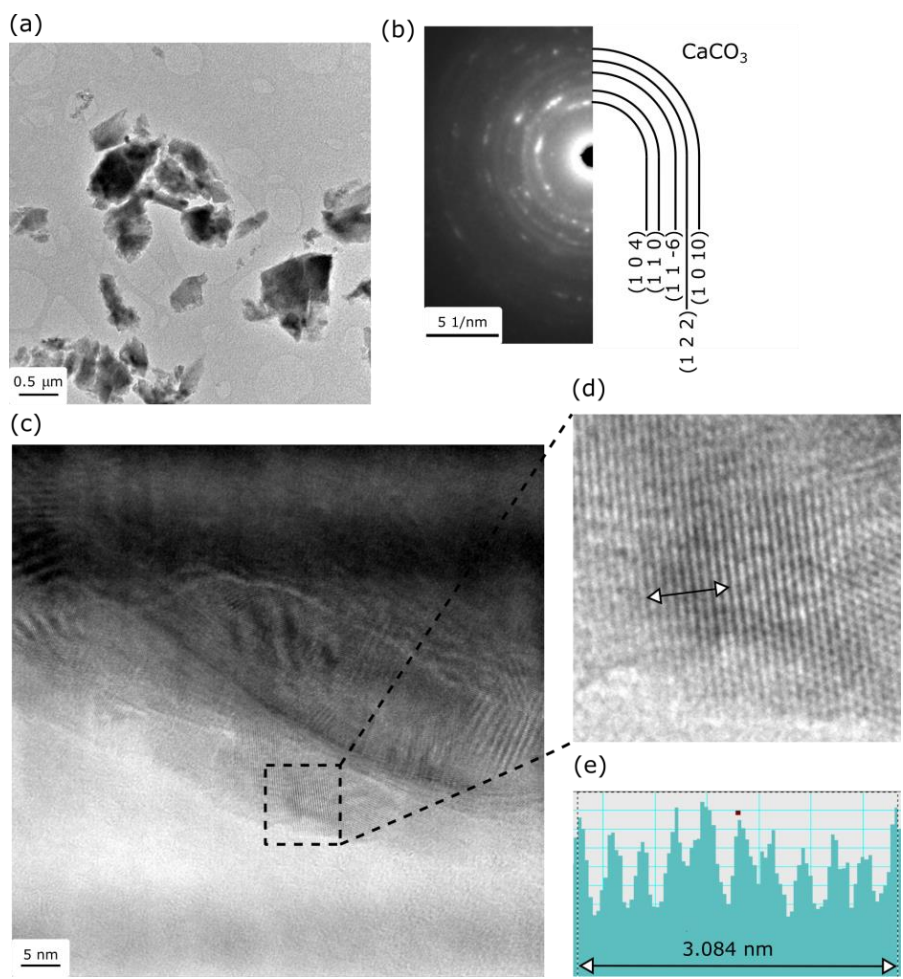

*Figure 4. TEM analysis of the fragments of precipitated particles: (a) Bright-field TEM image of the fragments of crushed precipitated calcite; (b) SAED obtained from the crushed precipitate shown in (a), the indexing of the diffraction rings matches well with the crystal structure of the calcite phase; (c) Higher-magnification TEM image of the precipitated particles, showing the presence of crystalline lattice fringes; (d) HRTEM image from the area marked with the black dashed square in (c); (e) Measurement of the interplanar spacing by averaging ten lattice fringes marked in (d).*

The experiments presented in this study have been repeated three times, confirming that the results are reproducible. However, it is still not clear what are the optimal conditions for fungal growth and $CaCO_3$ precipitation. Along with our previous work[45], we found $CaCO_3$ precipitates in both *A. nidulans* (MAD1445)-inoculated and *T. reesei* (ATCC13631)-inoculated media, but the amount of mineral crystals was rather limited. As our future endeavors, the effects of the multiple factors promoting fungal growth and $CaCO_3$ precipitation will be tested, such as concentrations of calcium ions and dissolved inorganic carbon, temperature, exposure to ultraviolet light, growth medium composition, fungal spore concentration, and availability of nucleation sites, etc. For examples, so far, our tests have confirmed that adding 4.0 µg/ml aminobenzoic acid and 0.02 µg/ml biotin to the growth medium significantly promotes the growth of *A. nidulans* (MAD1445); and adding 5.0 mM $CaCl_2$ to the growth medium can promote calcium carbonate precipitation. In addition, a systematic investigation will be conducted to assess the fungal spore germination. According to Wang et al., whether or not the fungal spores in/on the concrete crack germinate can be measured by the oxygen consumption on the crack surface[76]. Furthermore, we are also testing urease-positive fungi *Neurospora crassa*[77,78] and *Myrothecium gramineum*[79] for this specific application.

**4.5 Embedment of Fungal Healing Agents in Concrete**

If the fungal spores are larger than the pore sizes in concrete, if they are directly put into cement paste, most spores will be crushed and lost viability during the hydration process. Our results on μCT characterization of concrete microstructure have been shown in Fig. 5. The 3D renderings, as shown in Fig. 5(a) to Fig. 5(c), are particularly useful for qualitative assessment of phase distribution, which is complementary to higher resolution 2D microscopy imaging. From a typical 2D slice as shown in Fig. 5(d), it shows that the matrix pore sizes after 28 days of curing are much smaller than the typical diameters of fungal spores, which are usually larger than 3 μm (Fig. 1). The matrix pore sizes in air-entrained specimens after 28 days of curing are shown in Fig. 5(e). It shows that air-entraining agents could be applied to generate plentiful additional tiny air bubbles in concrete matrix to provide the housing for the healing agents. Although the specimen without air-entraining agents may also contain a small number of inhomogeneous air voids, the specimen containing air-entraining agents usually shows a large number of homogeneous air voids with appropriate size, which is more beneficial for both workability and durability of concrete.

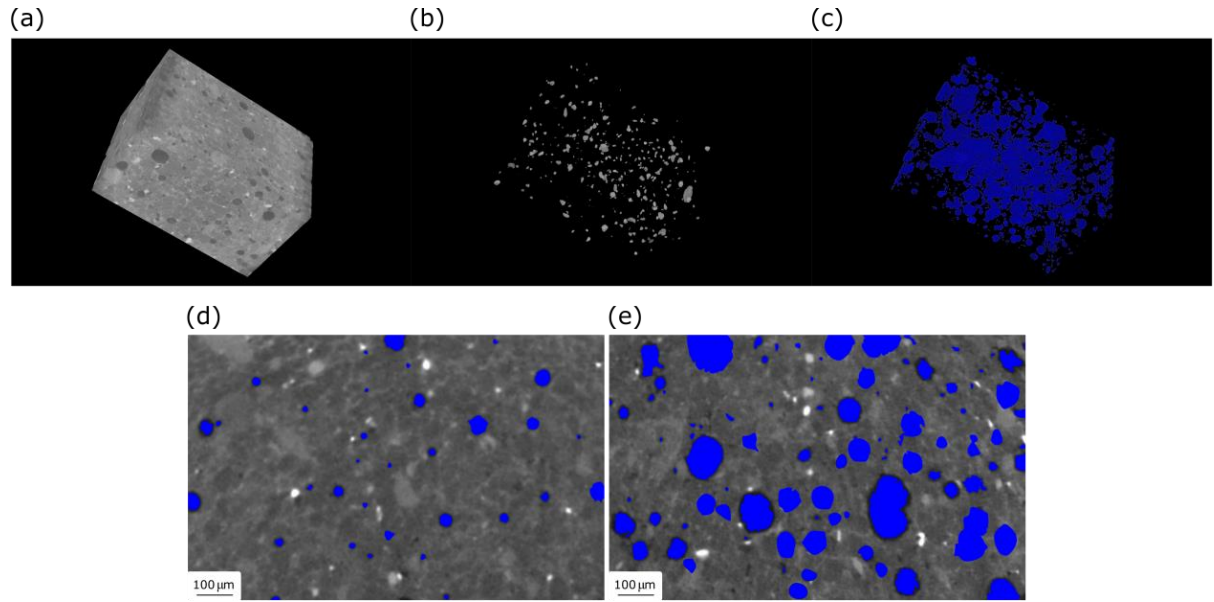

*Figure 5. A 3D rendering of a grayscale volume is shown in (a), the corresponding high-density particles and pores are shown in (b) and (c), respectively. (d) Pore size distribution inside cement paste specimens after 28 days of curing measured by μCT. (e) Effect of air-entraining agent on pore size distribution measured by μCT.*

**5. Concluding Remarks**

In this study, a novel self-healing approach is investigated, in which fungi are used to heal concrete cracks by promoting $CaCO_3$ precipitation. The experimental results showed that, despite the significant pH rise caused by the leaching of calcium hydroxide from concrete, one type of pH regulatory mutants of *A. nidulans* (MAD1445) could grow on concrete plates. In comparison, *A. nidulans* (ATCC38163), *A. nidulans* (MAD0305), *A. nidulans* (MAD0306), *R. oryzae* (ATCC22961), *P. chrysosporium* (ATCC24725), *A. terreus* (ATCC1012), and *A. oryzae* (ATCC1011) could not grow on concrete. The characterization by XRD, SEM, and TEM confirmed that the precipitates promoted by fungal activities were mainly composed of calcite.

It is important to note that according to the existing literature, *A. nidulans* is usually considered relatively harmless to healthy human beings. However, according to the investigation conducted by Bignnell et al.[80] on the role of the PacC-mediated pH adaptation in the pathogenesis of pulmonary aspergillosis, elimination of the PacC protein increases virulence in a murine neutropenic model, indicating that activities under PacC control may be crucial for *A. nidulans* pathogenicity. Therefore, a complete assessment needs to be performed to examine the possible short-term and long-term adverse outcomes of *pacC$^c$* mutation of *A. nidulans* on both environment and human beings before it is used in concrete structures.

The research findings from this study will also be useful to many other applications, such as removal of radionuclides, bio-consolidation of soil and sand, and removal of calcium ions from industrial wastewater. (1) The discharge of radionuclide wastewater leads to serious contamination in the environment[81]. Traditionally, chemical reactions, ion exchange, as well as membrane technologies are used to handle this challenge, but these methods are ineffective or costly. In contrast, microbial $CaCO_3$-based coprecipitation offers a cost-effective solution[82]. The current study will shed light on how fungi could effectively promote formation of radionuclide carbonate minerals. (2) Bio-consolidation is often involved in stabilization of erosion and increasing slope stability[83]. Microbe-mediated $CaCO_3$ precipitation offers a reliable and low-cost method to bind sand particles together, which can significantly influence the important mechanical properties of the sand, including permeability, compressibility, and shear strength[84]. The current study will help us understand how fungi could be used in such applications. (3) Large amounts of calcium ions in industrial wastewater often clog the pipelines, boilers, and heat exchangers[85]. Microbially induced $CaCO_3$ precipitation presents an effective and eco-friendly solution to remove inorganic contaminants from the environment. The current study will advance our knowledge in how fungi could applied to get rid of calcium ions from industrial wastewater.

**Acknowledgement**

Congrui Jin was funded by the Research Foundation for the State University of New York (TAE-16083068). Congrui Jin was also supported by the Small Scale Systems Integration and Packaging (S3IP) Center of Excellence. Dr. Miguel Penalva at the Biological Research Center of the Spanish National Research Council was thanked for kindly providing three different alkalinity-mimicking mutants of *A. nidulans,* i.e., MAD1445, MAD0306, and MAD0305. Dr. Joan Tilburn at Department of Microbiology of Imperial College London, Dr. Daniel Lucena-Agell at the Biological Research Center of the Spanish National Research Council, and Dr. America Hervas-Aguilar at the Division of Biomedical Sciences of University of Warwick were thanked for useful discussion.

**Author Contributions Statement**

CJ and NZ conceived and initiated the project. RRM, JL, HZ, and XC designed and performed the experimental work. GZ, NZ, and CJ supervised the research tasks. RRM and CJ wrote the initial draft of the manuscript. RRM, JL, XC, ZL, GZ, NZ, and CJ co-edited the manuscript. All authors reviewed the manuscript before submission.

**Competing Interests**

The authors declare that they have no competing interests.

**Data Availability Statement**

The datasets generated during and/or analyzed during the current study are available from the corresponding author on reasonable request.